\input phyzzx\def\e{\adveq\eqno{\rm (\chapterlabel\the\equanumber)}}
\def\mysec#1{\equanumber=0\chapter{#1}}
\def\adveq{\global\advance\equanumber by 1}
\def\myeq{{\rm \chapterlabel.\the\equanumber}}
\def\rarrow{\rightarrow}
\def\d#1{{\rm d}#1\,}

\def\twoline#1#2{\displaylines{\qquad#1\hfill(\adveq\myeq)\cr\hfill#2
\qquad\cr}}

\def\manyeq#1{\eqalign{#1}\e}

\def\semidirect{\mathrel{\raise0.04cm\hbox{${\scriptscriptstyle |\!}$
\hskip-0.175cm}\times}}

\def\mod{\mathop{\rm mod}\nolimits}

\def\ref#1{$^{[#1]}$}

\def\pr#1{#1^\prime}
 
\def\r#1{$[\rm#1]$}

\def\d{{\rm d}}

\def\Tr{\mathop{\rm Tr}\limits}
\def\e{\adveq\eqno{\rm (\chapterlabel.\the\equanumber)}}
\def\mysec#1{\equanumber=0\chapter{#1}}
\def\adveq{\global\advance\equanumber by 1}
\def\myeq{{\rm \chapterlabel.\the\equanumber}}
\def\rarrow{\rightarrow}
\def\d#1{{\rm d}#1\,}

\def\twoline#1#2{\displaylines{\qquad#1\hfill(\adveq\myeq)\cr\hfill#2
\qquad\cr}}

\def\manyeq#1{\eqalign{#1}\e}

\def\semidirect{\mathrel{\raise0.04cm\hbox{${\scriptscriptstyle |\!}$
\hskip-0.175cm}\times}}

\def\mod{\mathop{\rm mod}\nolimits}

\def\ref#1{$^{[#1]}$}

\def\pr#1{#1^\prime}
 
\def\r#1{$[\rm#1]$}

\def\d{{\rm d}}

\def\Tr{\mathop{\rm Tr}\limits}
\def\Im{\mathop{\rm Im}\nolimits}
\def\half{{1\over2}}

\def\opl{\mathop{\oplus}\limits}
\overfullrule=0pt

\date{June, 2011}
\titlepage
\title{Nonstandard Parafermions and String Compactification}
\author{Andrei Babichenko and Doron Gepner}
\line{\hfill Department of Particle Physics\hfill}
\line{\hfill Weizmann Institute\hfill}
\line{\hfill Rehovot 76100, Israel\hfill}
\abstract
Nonstandard parafermions are built and their central charges and dimensions are calculated. We then
construct new $N=2$ superconformal field theories by tensoring the parafermions with a free boson.
We study the spectrum and modular transformations of these theories.
Superstring and heterotic strings in four dimensions are then obtained by tensoring the new
superconformal field theories along with some minimal models. The generations and antigenerations are
studied. We give an example of the $1^2(5,7)$ theory which is shown to have three net generations. 
\endpage
\par
\mysec{Introduction.}

Supersymmetric heterotic string theories, in four dimensions, have the potential for a string 
theory description of the particle
spectrum. Here we describe new such string theories. We have an infinite number of models, which provide
a vast playground for exploring string phenomenology.

The construction is based on the generalization of Zamolodchikov and Fateev (ZF) standard 
parafermions \REF\ZF{A.B. Zamolodchikov and V.A. Fateev, Zh. Eksp. Theor. Fiz. 89 (1985) 380.}\r\ZF. 
We, first,  build the nonstandard parafermions and calculate the central charges and 
the dimensions in these theories. 
Precisely for the central charge we find we may construct 
N = 2 superconformal field theories, by adding one free boson at a certain radius.
This mimics the construction from standard parafermions of the minimal $N=2$ superconformal field theories
\REF\Ntwo{Z. Qiu, Phys. Lett. B 188 (1987) 207; A.B. Zamolodchikov and V.A. Fateev, Zh. Eksp. Theor. Fiz. 
90 (1986) 1553.}\r\Ntwo. 

Space time supersymmetric heterotic--like compactifications in four dimensions
can be achieved by means of an $N=2$ superconformal field theory on the world sheet
with the total central charge $9$. These are the so called $N=2$ string theories
\REF\Gep{D. Gepner, Nucl. Phys. B 296 (1988) 757.}
\r\Gep\ (for a review see 
\REF\Trieste{D. Gepner, Lectures in Trieste spring school, HEP (1989) 80.}\r\Trieste).
The original examples were constructed by tensoring the $N=2$ minimal models. We follow the, so called,
beta method to construct the superstring and heterotic string compactifications to any even dimension.
To get heterotic string we use the heterotic map \r\Gep.

This procedure results in full fledged and fully consistent string theories in four dimensions.
The theories have a gauge symmetry group which contains $E_8\times E_6$ along
with chiral fermions in the $27$ and $\bar{27}$ of $E_6$. Thus, these new theories are candidates to describe
the known particle spectrum. In section (5), we give an example of the theory $1^2(5,7)$ which is
shown to have four generations and one antigeneration, and is thus a good candidate for phenomenology. 

In ref. \REF\Geptwo{D. Gepner, Phys. Lett. B 199 (1987) 380.}\r\Geptwo\ a connection between $N=2$ string theory and
Calabi--Yau geometry was discussed. Compactifications on these manifolds were originally discussed
by Candelas et al. \REF\Cand{P. Candelas, G. Horowitz, A. Strominger and E. Witten, Nucl. Phys. B 258 (1985) 46.}
\r\Cand. In ref. \r\Geptwo\ it was conjectured that 
all $N=2$ string theories can be viewed as compactifications on some Calabi--Yau manifold.
For example, the theory $3^5$ (five copies of the third minimal model) was shown
to correspond to the quintic hypersurface in $CP^4$ \r\Geptwo.

Thus, it is an interesting question what manifolds correspond to the new string theories described here.
We address this question in the discussion.

\mysec{Nonstandard parafermions and $N=2$ superconformal field theory.}
 
The idea is to generalize the construction of the minimal $N=2$ superconformal field theory, using nonstandard parafermions. Specifically,
we have for the parafermions, $\psi_n$, $n=1,2,\ldots,N-1$, the following dimension formula,
$\Delta_n=\Delta_{N-n}$ and
$$\Delta_n=m n(N-n)/N+M_n.\e$$
The standard case corresponds to $m=1$ and $M_n=0$.

The parafermions are field $\psi_i(z)$ such that $i$ is defined modulo $N$ and
$\psi_i(z)$ carries the $Z_N$ charge $i$. We also assume $\psi_i^\dagger(z)=\psi_{N-i}(z)$. 

The parafermions obey the operator products as follows,
$$\psi_i(z)\psi_j(w)=c_{i,j} {\psi_{i+j}(w)\over (z-w)^{\Delta_i+\Delta_j-\Delta_{i+j}}},\e$$
up to higher order terms, and $c_{i,j}$ are some constants,
for $i+j\neq 0\mod N$. Also,
$$\psi_i(z) \psi_{N-i}(w)={1\over (z-w)^{2\Delta_i}}\left[1+(2\Delta_i/c)(z-w)^2 T_\psi(w)\right],\e$$
up to higher order terms. Here $c$ is the central charge of the theory and $T_\psi(w)$ is the stress tensor.
In the subsequent discussion we will specialize to the case where $1<\Delta_1< 3/2$ and
$$4\Delta_1-\Delta_2=2.\e$$
Other cases are relevant, but we shall not deal with them here, see, for example, 
\REF\Dot{V.S. Dotsenko and J.L. Jacobsen, R. Santachiara,
Nucl.Phys. B 679 (2004) 464;
V.S. Dotsenko, J.L. Jacobsen and R. Santachiara,
Nucl.Phys. B664 (2003) 477}\r\Dot.
We leave the rest of the dimensions, $\Delta_3,\Delta_4,\ldots, \Delta_{N-3}$, determined only up to some unknown
integers, which are presumably specified
by the unitarity requirement of the theory. 

We consider the $2r$ point function,
$$\left\langle \psi_1(z_1) \psi_1(z_2)\ldots\psi_1(z_r) \psi_1^\dagger(w_1)\psi_1^\dagger(w_2)\ldots 
\psi_1^\dagger (w_r) \right\rangle.\e$$
We define the rational negative number $M$ by
$$M={1\over 1-\Delta_1}.\e$$
The case of $M$ positive integer is the standard parafermions case.
If we multiply this correlation function by
$$\prod_{i=2}^r (z_1-z_r)^{2/M} \prod_{j=1}^r (z_1-w_j)^{-2/M},\e$$
we get an analytic function in the variable $z_1$. This correlation function, as a function of $z_1$,
is then determined by the residues
in the complex plane, and it vanishes as $z_1\rarrow\infty$. 

This leads to the following relation,
$$\manyeq{
&\left\langle  \psi_1(z_1) \ldots\psi_1(z_r) \psi_1^\dagger(w_1)\ldots 
\psi_1^\dagger (w_r) \right\rangle=\cr
&\prod_{i=2}^r (z_1-z_r)^{-2/M}  \prod_{j=1}^r (z_1-w_j)^{2/M} 
\sum_{k=1}^r \bigg\{
{1\over (z_1-w_k)^2}+\cr
&{2/M\over (z_1-w_k) }\bigg[\sum_{l=2}^r {1\over w_k-z_l}-\sum_{m=1\atop m\neq k}^r
{1\over w_k-w_m}\bigg]\bigg\}\cr
&\times 
\prod_{q=2}^r (z_q-w_k)^{2/M}\prod_{p=1}^{k-1} (w_p-w_k)^{-2/M} \prod_{s=k+1}^r (w_k-w_s)^{-2/M}\cr
&\left\langle \psi_1(z_2)\ldots \psi_1(z_r)\psi_1^\dagger(w_1)\ldots \psi_1^\dagger(w_{k-1})\psi_1^\dagger(w_{k+1})\ldots
\psi_1^\dagger(w_r)\right\rangle,\cr}$$
which expresses the $2r$ correlation function, in terms of the $2r-2$ correlation function.
In fact, this formula is identical to the one found by ZF, ref. \r\ZF, eq. (3.12) there, 
for the standard parafermions, 
where $M$ is a
positive integer. The only difference is that now $M$ is allowed to be a negative rational number.
The central charge of the theory is obtained by evaluating the four point function, and we find
$$c={2(M-1)\over M+2}={2\Delta_1\over 3-2\Delta_1}.\e$$
Again this is the same central charge of standard prafermions if we identify $M$ as the positive integer $N$.
In light of this analogy, we may speculate that the new parafermions are some sort of analytic continuation
of the standard parafermions.

Note that the condition eq. (2.4) is enough to determine all the correlators eq. (2.5),
for any rational value of the dimension $\Delta_1$. We choose $1<\Delta_1<3/2$ since this is the window 
of unitarity of the theory. For $\Delta_1>3/2$ the central charge eq. (2.9) is negative so clearly the 
theory cannot be unitary. For $\Delta_1<1/2$ either $\Delta_1$ or $\Delta_2$ are negative so, again, the
theory cannot be unitary. Finally, for $1/2\leq \Delta_1<1$ the theory can be completed to an $SU(2)$ affine theory,
in the standard way, whose level is $M$. Since $SU(2)$ is unitary only for positive integer level $M$, this
implies that the only unitary models, with $\Delta_1<1$, are the standard ZF parafermions.
Non--unitary models are also interesting. However, for our purpose, building string theories, unitary models are
required and so we impose the condition $1<\Delta_1<3/2$.

For this exact central charge, eq. (2.9), the parafermions can be completed to an
$N=2$ superconformal field theory by adding one boson, as in the standard case,
$$\eqalign{
G_{+}(z)&=\sqrt{2(c+1)\over 3}\psi_1(z) :e^{i\beta \phi(z)}:,\cr
G_-(z)&=\sqrt{2(c+1)\over 3}\psi_1(z)^\dagger :e^{-i\beta \phi(z)}:,\cr
J(z)&=i\beta^{-1}\partial\phi,\cr
T(z)&=T_\psi(z)-\half :(\partial\phi)^2:,\cr}\e$$
where $\phi(z)$ is a canonical free boson and we choose,
$\beta^2/2+\Delta_1=3/2$. Here $T(z)$ is the stress tensor, $J(z)$ is the $U(1)$ charge and $G_\pm(z)$ are the 
superpartners of the stress tensor.

To find the spectrum of the theory, we proceed as in the case of the standard parafermions. We suppose
that the dimension of the first parafermion is
$$\Delta_1=-{m\over N}\mod Z,\e$$
where $m$ and $N$ are strange integers and $1<\Delta_1<{3\over2}$.
We divide the Hilbert space according to the $Z_N\times {\bar Z}_N$
charges $\{k,l\}$,
$$\{F\}={1\over2} \opl_{N\geq q,\bar q\geq 1-N\atop q+\bar q\in 2Z} \{F_{q,\bar q}\},\e$$
where $[q,\bar q]=[k+l,k-l]$. The field $\rho_{k,\bar k}\in F_{k,\bar k}$ will then have the spin,
$$s_{k,\bar k}=-{m(k^2-\bar k^2)\over 4N}\mod Z,\e$$
and the mutual semi--locality exponent between the fields $\rho_{k,\bar k}$ and $\rho_{q,\bar q}$ will then be
$$\gamma^{q,\bar q}_{k,\bar k}=-m(kq-\bar k\bar q)/(2N),\e$$
which is the phase we have when circling one of the fields around the other.
The parafermions themselves $\psi_k$ are in the space $F_{2k,0}$. The spin fields $\sigma_k$, and its 
dual $\mu_k$ are in the
representations $F_{k,k}$ and $F_{k,-k}$, respectively.

We define the generators of the parafermionic algebra, through the OPE,
$$\psi_1(z) \rho_{k,\bar k}(0)=\sum_{r=-\infty}^\infty z^{-[mk/N]-f_k+r-1} A_{{1+k\over N}-r}\rho_{k,\bar k}(0),\e$$
and 
$$\psi_1^\dagger(z) \rho_{k,\bar k}(0)=\sum_{r=-\infty}^\infty z^{[mk/N]+{\hat f}_k+r-1} A_{{1-k\over N}-r}^\dagger
\rho_{k,\bar k}(0).\e$$  
$[x]$ is defined as the fractional part of $x$, $[x]=x\mod Z$ and $0<[x]\leq 1$, and $[0]=0$.
$f_k$ and ${\hat f}_k$ are some integers to be determined, $f_0={\hat f}_0=0$.
Together $A$ and $A^\dagger$ generate the parafermionic algebra. Note that for the sake of brevity, the
index of the operators $A_{{1+k\over N}-r}$ and $A_{{1-k\over N}-r}^\dagger$ do not correspond to the
increment of dimension when acting with the operator, and is just a notation for keeping track of the sector.
The dimension increment follows from eqs. (2.15,2.16). 

By considering the path integral
$$\oint{\d z_1\over 2\pi i} \oint{\d z_2\over 2\pi i} \psi_1(z_1) \psi_1^\dagger(z_2) z_1^{[km/N]+n+f_k}
z_2^{-[km/N]+r-{\hat f}_k} (z_1-z_2)^{-(3-2\Delta_1)} \rho_{k,\bar k}(0),\e$$
where the exponents were chosen to make the integral single valued,
we get the generalized commutation relation of the parafermionic operators $A$ and $A^\dagger$,
$$\twoline{\sum_{l=0}^\infty C^l_{-(3-2\Delta_1)} \bigg( A_{-{1-k\over N}+n-l-1} A^\dagger_{{1-k\over N}+r+l+1}-
A^\dagger_{-{k+1\over N}+r-l} A_{{1+k\over N}+n+l}\bigg)=}{
(3-2\Delta_1) L_{n+r+f_k-{\hat f}_k}+{1\over2} (n+f_k+[km/N])(n-1+f_k+[km/N])\delta_{n+r+f_k-{\hat f}_k,0},}$$
where we have the coefficients,
$$C^l_\lambda={\Gamma(l-\lambda)\over l!\Gamma(-\lambda)}.\e$$
Similarly we get the generalized commutation relations of the $A$'s with themselves,
$$\sum_{l=0}^\infty C^l_{2-2\Delta_1} \left(A_{(3+k)/N+n-l} A_{(1+k)/N+r+l}-A_{(3+k)/N+r-l} A_{(1+k)/N+n+l}\right)=0.\e$$
Similar result holds for the operators of the type $A^\dagger$.  

We define the spin field $\sigma_k$, $k=0,1,2,\ldots,N$, by the relation,
$$A_{(1+k)/N+n-f_k}\sigma_k=A^\dagger_{(1-k)/N+n+1+{\hat f}_k}\sigma_k=0,\e$$
where $n\geq 0$.
Similar relation holds for the right moving part.
By applying eq. (2.18), with $n=-f_k$ and $r={\hat f}_k$, we find the dimensions of the spin fields,
$$d_k={[m k/N] (1-[mk/N]) \over 2(3-2\Delta_1)}.\e$$

We define the rest of the primary fields by acting on $\sigma_k(z)$ with the parafermionic generators $A$
and $A^\dagger$. We have a series of fields $\phi^k_q(z)$ where $\phi^k_q=\phi^k_{q+2 N}$ and $q$ is defined
modulo $2N$, given by
$$\phi^k_{k+2l}=A_{(k-1+2l)/N-f_k-1} A_{(k-1+2l-2)/N-f_k-1} \ldots A_{(1+k)/N-f_k-1} \sigma_k,\e$$
where $l=0,1,\ldots ,N-k$.  
This field has the left dimension
$$\Delta_{k+2l}^k=d_k+m{l(N-k-l)\over N},\e$$
and
$$\phi^k_{k-2 l}=A^\dagger_{-(1+k-2l)/N+{\hat f}_k} A^\dagger_{-(1+k+2-2l)/N+{\hat f}_k} \ldots 
A^\dagger_{(1-k)/N+{\hat f}_k} \sigma_k,\e$$
where $l=0,1,\ldots,k$. The left dimension is
$$\Delta^k_{k-2l}=d_k+m{l(k-l)\over N}.\e$$
The dimensions are defined up to an integer which we do not know, except for the spin and dual spin fields
whose dimensions are known exactly, eq. (2.22).  
Similar fields are obtain by acting with the right moving parafermions. In the sequel we will denote the two fields
eq. (2.23,2.25) as $\phi^k_l$ where we redefined $l$ instead of $k\pm 2l$. Of course, the field is defined only for
$k+l=0\mod 2$ and is assumed as zero otherwise. The integers $f_k$ are such that $\Delta_1+\Delta_k=\Delta_{k+1}+
[2km/N]+f_{2k}$, where $\Delta_k$ is the dimension of the parafermion $\psi_k$. Analogous relation holds for
${\hat f}_k$. 

Let us turn now to the $N=2$ superconformal field theory, where the algebra was defined by eq. (2.10).
The general primary field of the theory has the form,
$$\Lambda^k_q(z)=\phi^k_q(z) :e^{i\gamma_q\phi(z)}:,\e$$
where $\gamma_q$ is given from locality with $G_\pm(z)$, and is given in the NS sector by,
$$\gamma_q={1\over \beta} \left[ {mq\over N}\right],\e$$
and it has the $U(1)$ charge
$$Q^k_q={1\over \beta^2} \left[ {mq\over N}\right],\e$$ 
and the dimension,
$$D^k_q=\gamma_q^2/2+\Delta^k_q.\e$$

The chiral fields are the fields which obey $D^k_q=Q^k_q/2$, and it is easy to see 
that these are precisely the fields obtained when $q=k$ and they have the 
dimensions and $U(1)$ charges,
$$Q^k_k={c+1\over 3} \left[ {mk\over N}\right],\e$$
$$D^k_k={c+1\over 6} \left[ {mk\over N}\right],\e$$
and $k=0,1,2,\ldots,N$.
We see that we have exactly $N+1$ chiral fields, which is the Witten index of the theory, 
and the top one has the $U(1)$ charge $(1+c)/3$, as it should.
This indicates, although we have not proven this, that the theory is unitary, as we seem to get a 
consistent unitary $N=2$ superconformal field theory. 	

\mysec{The characters and the partition functions.}
\def\twiddle#1{{\hat #1}}
\def\pr#1{{{#1}^\prime}}
\par
We denote the primary fields of the parafermionic theory as
$$\Phi^k_l(z),\e$$
in view of eqs. (2.23,2.25), $k+l=0\mod 2$, $k=0,1,2,\ldots, N$ and $l$ is defined modulo $2N$. 
The dimension of this field is
$$\Delta^k_l=d_k+{m k^2\over 4N}-{m l^2\over 4 N},\e$$
up to an integer,
where $d_k$ was given by eq. (2.22). We define the character of the representations
of the parafermionic algebra by
$$c^k_l(\tau)=\Tr_{{\cal H}^k_l} e^{2\pi i \tau(L_0-c/24)} ,\e$$
where $L_0$ is the dimension operator and the Hilbert space ${\cal H}^k_l$ is the
space of all the descendants by integer dimensions products of the $A$ operators on the 
field $\Phi^k_l$.
In light of the dimension formula eq. (3.2), the following relation is suggested for the 
characters,
$$\sum_{l\mod 2 N} c^k_l(\tau) \Theta^{(m)}_{l,N}(\tau)/\eta(\tau)^f=B^k(\tau),\e$$
where $\Theta^{(m)}_{l,N}(\tau)$ is a level $N$ $SU(2)$ theta function at pseudo level $m$.
Strictly speaking, this theta function is defined only for $m$ which is odd.
When $m$ is even, a different theta function has to be chosen. Thus, in the sequel, we shall
assume that $m$ is odd. We believe, however, that the subsequent results hold also for even $m$,
with some necessary changes.

The modular transformations of this theta function are
$$\Theta^{(m)}_{l,N}(\tau+1)=e^{2\pi i\Delta_l} \Theta^{(m)}_{l,N}(\tau),\e$$
where 
$$\Delta_l={ml^2\over 4N},\e$$
and 
$$\Theta^{(m)}_{l,N}(-1/\tau)=(-i\tau)^{f/2} \sum_{r\mod 2N} S_{r,l}\Theta^{(m)}_{r,N}(\tau),\e$$
where the matrix $S_{l,r}$ is given by
$$S_{l,r}={1\over \sqrt{2N}} e^{-\pi i m lr/N},\e$$
and $f$ is the dimension of the bosonic lattice (equal to the central charge of the bosons).
The actual character is obtained by dividing the theta function, $\Theta^{(m)}_{l,N}$ by $\eta(\tau)^f$, which is the 
contribution to the partition function of the bosonic creation operators. 
The $B_k$ are conjectured to be the characters of some unknown theory.
We know just the dimensions of the characters $B_k$ which are 
$$b_k=d_k+{m k^2\over 4N},\e$$
up to an integer. We propose that the $B_k$'s form a representation of the modular
group,
$$B^k(-1/\tau)=\sum_{k,p} W_{k,p} B^k(\tau),\e$$
where $W_{k,p}$ is some unitary and symmetric matrix, as yet unknown. 
In what follows, we shall not require 
the explicit form of $W_{k,p}$.
The theta functions $\Theta^{(m)}_{l,N}$ can be realized as ordinary theta functions on
some lattice. According to ref. \REF\Baver{E. Baver, D. Gepner and U. Gursoy, Nucl. Phys. B 557 (1999) 505.}\r\Baver\ 
such a realization of $\Theta^{(m)}$ as an ordinary theta function on some lattice is
always possible. However, many lattices give the same modular 
transformations and we do not know the one suitable in eq. (3.4). In the sequel,
for the purpose of building the string theory partition function,
we shall require only the modular transformations of $\chi^k_l$,
which are given, in light of eq. (3.4) by
$$c^k_l(\tau+1)=e^{2\pi i(\Delta^k_l-c/24)}c_{k,l}(\tau),\e$$
$$c^k_l(-1/\tau)={1\over\sqrt{2N}}\sum_{p=0}^N\, \sum_{r\mod 2N} W_{p,k} e^{\pi im r l/N} c^p_r(\tau).\e$$
  
A modular invariant for the parafermionic system is 
defined by a positive integer matrix $M^{k,p}_{l,r}$ such that the partition function
$$Z=\sum_{k,p=0}^N\, \sum_{l,r\mod 2N} M^{k,p}_{l,r} c^k_l(\tau) {c^p_r}(\tau)^\dagger,\e$$
is modular invariant. Given a modular invariant of the $B^k$ system,
denoted by the positive matrix $N^{k,p}$  
and a modular invariant of the level $N$ theta functions, $D_{l,r}$ (which were
all classified in ref. \REF\GepQiu{D. Gepner and Z. Qiu, Nucl. Phys. B 285 [FS19] (1987) 423.}
\r\GepQiu) we can build the parafermionic invariant, the array $M$, as following,
$$M^{k,p}_{l,r}=\half N^{k,p} D_{l,r},\e$$
where the half accounts for the field identifications. 
Namely, the fields in the character $c^k_l(\tau) c^p_r(\bar\tau)$ are the same fields when we take,
$k\rarrow N-k$, $p\rarrow N-p$, $l\rarrow l+N$ and $r\rarrow N+r$. This is consistent with the
formula for the dimensions, eq. (3.2).
 
The only known invariant 
for the $B^k$ system, at the present, is the diagonal one,
$$N^{k,p}=\delta_{k,p}.\e$$

Let us turn now to the $N=2$ superconformal field theory built out of these parafermionic 
field theories. We wish to compute the characters of the representations. It is convenient to consider
the action of $G_\pm^2$ in the construction of the representations. We then have four sectors
$s=0,2$ for the NS sector and $s=1,3$ for the Ramond sector, where $s$ is defined modulo $4$.
We denote the character accordingly as
$$\chi^{k (s)}_q(\tau)\e$$
where $k$ and $q$ label the parafermionic field $\phi^k_q$ tensored with some bosonic operator.
The $U(1)$ charge, up to an even integer, and the dimension, up to an integer, are then all the same 
inside each such representation and are given by,
$$J^{k(s)}_q={1\over \beta^2} [mq/N]-s/2,\e$$
is the $U(1)$ charge, and the dimension is,
$$\Delta^{k(s)}_q=d_k+{m k^2\over4N}-{m(q-s)^2\over4N}+{1\over 2} \left({1\over\beta}[mq/N]-{s\beta\over 2}\right)^2
.\e$$
Here we demand $k+q+s=0\mod2$.

To get the characters $\chi^{k(s)}_q(\tau)$ we note that acting with $G_\pm^{2N}$ leaves us within the same
representation, or we are allowed a shift in the bosonic momenta of $p\rarrow p+2 N\beta$, inside each representation. 
This implies 
that the characters are expressed in terms of level $l=2N^2\beta^2$ $SU(2)$ classical theta functions, defined by
$$\theta_{n,l}(\tau)=\sum_{j={n\over 2l} \mod Z} e^{2\pi i \tau l j^2}.\e$$
The modular transformations of the theta function are given by eqs. (3.5,3.7) when substituting $m=1$,
$$\theta_{n,l}(\tau)=\Theta^{(1)}_{n,l}(\tau).\e$$
We can also act with $G_+^2$ which gives a simultaneous shift of $q\rarrow q+4$ and for the momenta
by $p\rarrow p+2\beta$. All this implies that the characters are given by the following important
formula,
$$\chi^{k(s)}_q={1\over\eta(\tau)}\sum_{j \mod N} c^k_{q+4j-s}(\tau) \theta_{2 s_q+N\beta^2(4j-s),2N^2\beta^2}(\tau),\e$$
where $s_q$ is an integer defined by
$$s_q=N[mq/N].\e$$

We have for the dimension of the first parafermion, $\Delta_1$,
$$\Delta_1=2-{m\over N},\e$$
where $m$ is an integer strange to $N$, such that $N/2<m<N$.
Then $\beta$ is 
$$\beta^2={2m-N\over N},\e$$
and the central charge of the theory is 
$$\hat c=c+1={3N\over 2m-N}.\e$$
The level of the theta function is thus,
$$l=2N^2\beta^2=2N(2m-N),\e$$
and $N\beta^2=2m-N$. The $U(1)$ charge of the fields in the character $\chi^{k(s)}_q$ then assumes the form
$$J^{k(s)}_q=-{s\over2}+{N\over 2m-N} [mq/N],\e$$
up to an even integer.

From the character formula, eq. (3.21), we find the dimensions of the fields appearing in the character. 
These are 
$$d_k+{m k^2\over 4N}-{m(q+4j-s)^2\over 4 N}+{(2 s_q+N\beta^2 (4j-s))^2\over 8 N^2\beta^2}.\e$$
Remarkably, this dimension can be written up to an integer as
$$d_k+{m k^2\over 4N}-{(s+2 r_q)^2\over 8}+{m(q+2 r_q)^2\over 4(2m-N)},\e$$
where $r_q$ is an integer defined by
$$[mq/N]=mq/N+r_q.\e$$
This implies that the characters can be expressed, as far as the dimensions are concerned, as
$$\sum_{s\mod4} \chi^{k(s)}_q(\tau)\theta_{s+2r_q,2}(\tau)=B^k \Theta_{q+2 r_q,2m-N}^{(m)}(\tau)/\eta(\tau)^{f-1},\e$$
where the pseudo theta function, $\Theta^{(m)}$, was defined, through its modular transformations, 
by eqs. (3.5,3.7). This also implies, because of the appearance of the theta functions
in eq. (3.31), that the theory has the discrete
symmetry $Z_2\times Z_{2m-N}$, which will be of importance when discussing the string theory compactifications. 
The $U(1)$ charge, up to an even integer, is written in terms of these variables as
$$J^{k(s)}_q=-{s+2r_q\over 2}+{m(q+2 r_q)\over 2m-N}.\e$$

To substantiate this character relation, eq. (3.21), we need to verify, also, the behavior under the modular 
transformation $S:\tau\rarrow -1/\tau$. However, since the $q$ and $s$ variables transform under modular
transformations as theta functions,
which are essentially free massless bosons, and the matrix $S$ can be written as $\exp[-2\pi i (\Delta_{\lambda+\mu}
-\Delta_\lambda-\Delta_\mu)]$, where $\Delta$ is the dimension, for some vectors $\lambda$ and $\mu$, it is
enough to check
the dimensions and then invariance under $S$ will follow from that. 
We conclude that as far as modular transformations are concerned
the character relation eq. (3.31) holds. We emphasize that we do not know the appropriate realization for the
theta function at pseudo level $m$, $\Theta^{(m)}$. 
Namely, $\Theta^{(m)}$ is an ordinary theta function on some lattice, which we do not know.
In the sequel, however, we will only require the modular 
transformations, since our focus will be on building the string theory partition function, along
with analyzing its massless spectrum.

It is easy now to classify the modular invariants of these $N=2$ superconformal field theories.
To do this define the `shifted' character,
$${\twiddle \chi}^{k(s)}_q(\tau)=\chi^{k(s+2 r_q)}_{q+2 r_q}(\tau).\e$$
A modular invariant of the theory is obtained by requiring the modular
invariance of the partition function
$$\sum_{k,\pr k=0}^N\, \sum_{s,\pr s \mod 4}\, \sum_{q,\pr q\mod 4m-2N} R^{k (s) \pr k (\pr s)}_{q \pr q} {\twiddle \chi}^{k(s)}_q(\tau) 
{\twiddle \chi}^{\pr k (\pr s)}_{\pr q}(\tau)^\dagger,\e$$
where $ R^{k (s) \pr k (\pr s)}_{q \pr q}$ is an array of non--negative integers.
We can then write a solution for the multiplicities array, $R$, in terms of the theta function invariants,
$$ R^{k (s) \pr k (\pr s)}_{q \pr q}=\half N_{k,\pr k} S^2_{s,\pr s} S^{2m-N}_{q,\pr q},\e$$
where $N_{k,\pr k}$ is a modular invariant of the $B^k$ field theory, and $S^l_{q,\pr q}$ denotes the
modular invariants of the theta function at the level $l$. The latter were all classified in ref. \r\GepQiu.
In special cases there might be additional invariants such as orbifolds of the combined discreet symmetries
$Z_2\times Z_{2m-N}$. The factor $\half$ is for the presumed field identifications. Namely, we get the same field when
we take 
$k\rarrow N-k$, $q\rarrow q+2m-N$ and $s\rarrow s+2$, simultaneously on both the right and left movers. This
is consistent with the dimension formula, eq. (3.29).

\mysec{New string compactifications.}
We turn now to building the string theory partition function. We assume that we have several models of the new
parafermionic $N=2$ superconformal field theories, denoted by the two strange integers, $m_i$ and $N_i$, $i=1,2,\ldots n$,
where $m_i$ obeys, $N_i/2<m_i<N_i$. The dimension of the first parafermion is $\Delta_1^i=2-m_i/N_i$. The 
central charges of the $N=2$ superconformal field theories are $c_i=3N_i/(2m_i-N_i)$. We also allow to tensor
along with these theories minimal $N=2$ superconformal field theories, labeled by the integers $k_i$ where $i=1,2,\ldots,r$.
We wish to build a consistent string compactification to $D$ dimensions, where of main interest is the case 
of four dimensions, $D=4$. The number of transverse dimensions is $d=D-2$, in the light cone gauge. 
Then the condition for the 
central charge becomes,
$$12={3d\over2}+\sum_{i=1}^r {3k_i\over k_i+2}+\sum_{i=1}^n {3N_i\over 2m_i-N_i}.\e$$

To build this theory we follow the method of ref. \r\Gep. We employ the so called beta method to get space time
supersymmetric $D$ dimensional superstring theory, where $D$ is assumed to be even. The characters of the minimal 
models are denoted by $\psi^{l_i(s_i)}_{m_i}(\tau)$ where $0\leq l_i\leq k_i$ and $m_i$ is defined modulo $2(k_i+2)$
and $s$, modulo $4$, labels the sector, $m_i+l_i+s_i=0\mod2$. The character $\psi$ obeys the formula
$$\sum_{f_i\mod 2(k_i+2)} \psi^{l_i (s_i)}_{f_i}(\tau) \theta_{f_i,k_i+2}(\tau)=A^{l_i,k_i}(\tau) \theta_{s_i,2}(\tau),\e$$
where the $\theta_{l,n}(\tau)$ denotes the level $n$ $SU(2)$ classical theta function, defined by eq. (3.19),
and $A^{l_i,k_i}(\tau)$, $l_i=0,1,\ldots,k_i$ is the character of the representation with isospin $l_i/2$ of 
the affine $SU(2)$ WZW theory, at the level $k_i$. From the eq. (4.2) it follows that under modular transformations
the indices $s_i$ behave like level $2$ theta functions and the indices $f_i$ behave like level $-(k_i+2)$ 
theta functions. 

The characters of the new $N=2$ superconformal field theories are denoted by ${\hat\chi}^{l_i(s_i)}_{f_i}(\tau)$,
where $i=1,2,\ldots,n$. From, eq.(3.31) it follows that the indices $s_i$ behave like level $-2$ theta functions,
under modular transformations,
whereas the indices $f_i$ behave like level $(2 m_i-N_i)/m_i$ theta functions.
To these indices, we also tensor the world sheet fermions degrees of freedom, expressed by an $SO(d)$, level $1$
theta function $\theta_{\lambda,{\rm SO(d)}}$,  
$$\theta_{\lambda,{\rm SO(d)}}(\tau)=\sum_{\mu\in \lambda+M} e^{\pi i \tau\mu^2},\e$$
where $M$ is the root lattice of $SO(d)$ and
the representation $\lambda$ is, in the NS sector: $0$ for the singlet, $v$ for the vector representation;
in the $R$ sector: $s$ and $\bar s$ for the spinor and antispinor representations.

Let us describe the beta method. Denote by 
$$Z_{\vec v,\vec n}=\prod_{i=1}^n \theta_{v_i,n_i}(\tau),\e$$
a product of the theta functions. We define the scalar product by,
$$ \vec v\cdot \vec u=\sum_{i=1}^n {v_i u_i\over 2 n_i},\e$$
where $n_i$ can be any rational number (formally). We choose a vector $\beta$ such that 
$\beta\cdot \beta$ is an odd integer. A fermionic generalized character is then defined by
$${\hat Z}_{\vec v}=\sum_{b\in Z} (-1)^b Z_{\vec v+b\vec \beta,\vec n}(\tau),\e$$
provided that $\vec\beta\cdot \vec v+1/2\in Z$.
${\hat Z}_{\vec v}$ then forms a unitary representation of the modular group. We get a modular invariant 
partition function by multiplying
$$\sum_{\vec\beta\cdot\vec v+1/2\in Z} {\hat Z}_{\vec v}(\tau) {\hat Z}_{\vec v}(\tau)^\dagger.\e$$
We can also add $g$ bosonic beta vectors $\vec\beta_p$, where $p$ is $1,2,\ldots,g$, 
such that $\vec\beta_p\cdot \vec\beta_p$ is an even integer.
We also demand $\vec\beta_p\cdot\vec\beta_q$ and $\vec\beta\cdot\beta_p$ are all integers. Denote by $Q$ the lattice
generated by $\beta$ and $\beta_p$. Then the `generalized' character is
$$\hat Z_v(\tau)=\sum_{\vec u\in Q}
Z_{\vec v+\vec u},\e$$
provided that $v$ obeys
$$\vec\beta\cdot \vec v+1/2\in Z,\qquad {\rm and\ } \vec\beta_p\cdot\vec v\in Z.\e$$
will form a unitary representation of the modular group transforming by
$$\hat Z_{\vec v}(\tau+1)= e^{\pi i \vec v\cdot\vec v} \hat Z_{\vec v}(\tau),\e$$
Under $S:\tau\rarrow-1/\tau$ the generalized character transforms
by
$${\hat Z}_{\vec v}(-1/\tau)=C(-i\tau)^{f/2}\sum_{\vec u} e^{-2\pi i \vec v\cdot\vec u} {\hat Z}_{\vec u}(\tau),\e$$
where $f$ is the central charge of the bosonic system and 
$C$ is some constant determined by unitarity. $\vec u$ obeys the same integrality condition, eq. (4.9).
It means that ${\hat Z}_{\vec u}(\tau)$ forms a unitary representation of the modular group (when divided by 
$\eta(\tau)^f$). Thus, we get a modular invariant partition function
by the sum,
$$\hat Z=|\eta(\tau)|^{-2f}\sum_{\vec v\atop {\vec\beta\cdot\vec v+1/2\in Z\atop \vec\beta_p\cdot\vec v\in Z}}
{\hat Z}_{\vec v}(\tau) {\hat Z}_{\vec v}(\tau)^\dagger.\e$$

Let us return now to the string theory partition function. We have a tensor product of the $d$ transverse superstring
fermions, along with $r$ minimal models and $n$ new $N=2$ superconformal field theories. We assume that the central
charge condition, eq. (4.1), holds. We define the generalized character of this system by
$$Z^{\vec l}_{\vec v}=\theta_{v_1,SO(d)} \prod_{i=1}^{r} \psi^{l_i(v_{i+r+1})}_{v_{i+1}} \prod_{i=1}^n 
{\hat \chi}^{l_{i+r}(v_{i+2r+1+n})}_{v_{i+2r+1}},\e$$
where $\vec l$ is an $n+r$ long vector carrying the `affine' indices, and $\vec v$ is an $1+2(n+r)$ long vector
carrying the theta function indices, $\psi$ denotes the characters of the $r$ minimal models,
and $\hat\chi$ denotes the shifted characters of the $n$ nonstandard superconformal models.

We also define a modular invariant for the `affine' system. This is a product of $n$ level $k_i$ $SU(2)$ modular
invariants, denoted by $B_{l_i,{\bar l}_i}$, where ${l_i,\bar l}_i$ are integers obeying $0\leq l_i,{\bar l}_i\leq k_i$.
The invariants for the $SU(2)$ affine system are labeled by ADE refs. 
\REF\Cap{A. Cappelli, C. Itzykson and J.B. Zuber, Nucl. Phys. B 280 [FS16] (1987) 455.}\r{\GepQiu,\Cap}.
We also take an invariant for the 
new superconformal models $A_{l_{i+r},{\bar l}_{i+r}}$, where $i=1,2,\ldots,n$, for the `$B$ theory' indices.
The only known invariant, at the present, for the $B$ theories is the diagonal one,
$$A_{k,\bar k}=\delta_{k,\bar k}.\e$$  

We now employ the beta method to get a space time supersymmetric string theory in $D=d+2$ dimensions.
The scalar product in this case becomes,
$$\twoline{\vec v\cdot\vec u={v_1\cdot u_1}+\sum_{i=1}^{r} -{v_{i+1} u_{i+1}\over 2(k_i+2)}+\sum_{i=1}^r
{v_{i+r+1}u_{i+r+1}\over 4}+}{\sum_{i=1}^n {m_i v_{i+2r+1} u_{i+2r+1}\over 2(2m_i-N_i)}+
\sum_{i=1}^n -{v_{i+2r+n+1} u_{i+2r+n+1}\over 4}.}$$
For the vector $\vec\beta$ we take $\beta_1=s$, the spinor representation (for the transverse fermions) and
$$\beta_i=-1,\qquad i=2,3,\ldots, r+1,\e$$
and is $1$ everywhere else.
We can calculate 
$$\vec\beta\cdot\vec\beta={d\over 8}+\sum_{i=1}^r \left( -{1\over 2(k_i+2)}+{1\over 4}\right) +\sum_{i=1}^n
\left({m_i\over 2(2 m_i-N_i)}-{1\over 4}\right)={c_{\rm tot}\over 12}=1,\e$$
where $c_{\rm tot}$ is the total central charge. We conclude that the vector $\vec \beta$ has length
one as required from the fermionic beta method. Moreover, the condition, eq. (4.9), becomes for this beta
$\vec \beta\vec v+1/2\in Z$. It can be seen that $\vec\beta\vec v$ is half the total $U(1)$ charge
of the representation $Z_{\vec v}^{\vec l}$. This is since the total $U(1)$ charge is given by
$$\twoline{J^{\rm total}_{\vec v}=2\beta\cdot\vec v={2s\cdot v_1}+\sum_{i=1}^r {v_{i+1}\over k_i+2}+\sum_{i=1}^r
{v_{i+r+1}\over 2}+}{\sum_{i=1}^n{m_i v_{i+2r+1}\over 2m_i-N_i}-
\sum_{i=1}^n {v_{i+2r+n+1}\over 2},}$$
up to an even integer and according to eq. (3.32). 
This means that the total $U(1)$ charge is guaranteed to be an odd integer. This is the hallmark of
space time supersymmetry for the superstring partition function.
 
To the vector $\vec\beta$, we need to add $g$ bosonic beta vectors, $\beta_p$, $p=1,2,\ldots,g$, 
where $g$ is the number of level $\pm 2$
theta functions. We take $\vec \beta_p$ to be, vector on the $SO(d)$ theta function, $2$ on the $p$'th 
level $\pm2$ theta function and zero elsewhere. It is easy to see that $\vec\beta_p^2$ is an even integer,
and that $\vec\beta_p\vec\beta_q$ and $\vec\beta_p\vec\beta$ are all integers, for any $p$ and $q$. The condition of the beta
method on the representation $Z^{\vec l}_{\vec v}$ then reads: $\beta_p\vec v\in Z$. This implies
that $v_1\cdot s=v_t \mod 2$ for all the level $\pm2$ theta functions, denoted by $t$. This exactly ensures 
that states will be allowed only if the subtheories
are all in the NS sector or all are in the R sector. This ensures Lorenz invariance of the string theory.

We denote the lattice generated by $\beta$ and $\beta_p$ by $Q$. The superstring partition function is then given by the
beta method,
$$Z_{\rm superstring}=\sum_{\vec v\atop {\vec u\in Q\atop \vec v\beta+1/2\in Z\ \ \vec v\beta_p\in Z}}
\pm A_{\vec l,\vec{\bar l}} Z^{\vec l}_{\vec v} Z_{\vec v+\vec u}^{\vec{\bar l}},\e$$
where $A_{\vec l,\vec{\bar l}}$ is the affine modular invariant, where we take any of the ADE solutions
for the $SU(2)$ part, $l_i$ where $i=1,2\ldots,r$ and the diagonal modular invariant for the $B$ indices,
$l_i$ where $i=r+1,r+2,\ldots n+r$. We ommited the contribution
of the transverse world sheet bosons. The sign in the partition function is `$+$' for space time bosons and
`$-$' for space time spinors, as required by spin and statistics.  

To get an heterotic like string theory in $D$ dimension we use the so called heterotic map, which takes us from
a superstring theory to an heterotic one, ref. \r\Gep. The gauge group is chosen to be either
$G=SO(d+24)$ or $G=E_8\times SO(8+d)$. Denote by $Z_{\lambda,\bar\lambda}$ the partition function of
the internal theory which couples to the $\lambda$ ($\bar\lambda$) left $SO(d)$ representation (right $SO(d)$
representation), which are the left and right world sheet fermions representations in the superstring. 
Denote by $\theta_{\lambda,G}$ the four theta functions of the group $G$, ordered as
$0,v,s,\bar s$ (singlet,vector,spinor, antispinor). Then the matrix which implements the heterotic map
sends the singlet (vector) of $SO(d)$ to the vector (singlet) of $G$, and changes the signs of the spinors of $G$. 
It is given by
$$M=\pmatrix{0&1&0&0\cr 
             1&0&0&0\cr
             0&0&-1&0\cr
             0&0&0&-1\cr},\e$$
Then the heterotic string partition function is given by
$$Z_{\rm het}=|(\Im \tau)^{-d/2} \eta(\tau)^{-2d}| \sum_{\lambda,\bar\lambda} \theta_{\lambda,{\rm SO(d)}}(\tau)
(M\theta_G)_{\bar\lambda}^\dagger(\bar\tau) Z_{\lambda,\bar\lambda}(\tau,\bar\tau).\e$$
This partition function obeys spin and statistics and  the condition for the left and right central charges,
which is $24$ for the right movers (bosonic string) and $12$ for the left movers (fermionic string). 
This is a space time supersymmetric heterotic--like partition function in $D$ space--time dimensions.
The gauge group contains in four dimensions, $D=4$, the groups $E_8\times E_6$ or $SO(26)$.
Thus, we arrive at new string theories, which are fully consistent, full fledged theories.

\mysec{On the massless spectrum.}
Let us turn now to the massless spectrum of these new string theories in four dimensions. 
We have generations which are space time (left) chiral fermions
in the representation $27$ of $E_6$, along with anti--generations which are left moving fermions in the
representation $\bar{27}$ of $E_6$. In all $N=2$ string theories, the generations correspond to fields
which are left and right chiral (denoted by $(C,C)$) which have total $U(1)$ charge $1$ and dimension $\half$,
on both the left and right moving parts of the internal conformal field theories. The antigenerations
correspond, likewise, to left chiral field and right anti--chiral (denoted as $(C,A)$),
which have dimension $\half$ and $U(1)$ charge $-1$ for the right moving part. Thus, we need to study
the chiral fields in each of the subtheories.

For the $k$'th $N=2$ minimal superconformal model the chiral field is the lowest dimensional field in the
character,
$$\psi^{p(0)}_p(\tau),\e$$
where the character $\psi$ was defined in eq. (4.2), and $p=0,1,2,\ldots,k$. The $U(1)$ charge of this field is
$$J_p^k={p\over k+2},\e$$
and the dimension is half the $U(1)$ charge. The anti--chiral field is the dagger field, and it has the
same dimension and opposite $U(1)$ charge. The $(C,C)$ fields of this subtheory is an element of the
representation $\psi^{p(0}_p(\tau) \psi^{p(0)}_p(\tau)^\dagger$ and we denote them as
$$\Phi_p^k,\qquad p=0,1,2,\ldots,k,\qquad J_L=J_R={p\over k+2}.\e$$   
We assumed the diagonal modular invariant for the minimal model. Then there are no fields of the type $(C,A)$,
in the model. 

The chiral fields of the $(m,N)$ nonstandard $N=2$ superconformal field theories were given by eqs. (2.31,2.32). These
are the fields which are the lowest dimensional in the character $\chi^{q(0)}_q$, $q=0,1,2,\ldots, N$,
and their $U(1)$ charges are, eq. (2.31),
$$J^{m,N}_q={N\over 2m-N} \left[{mq\over N}\right],\e$$
and their dimensions is half the $U(1)$ charge. Again, we assume the diagonal modular invariant for this field.
The $(C,C)$ fields of the theory belong to the character $\chi^{q(0)}_q(\tau) \chi^{q(0)}_q(\bar\tau)$,
$q=0,1,2,\ldots,N$, and are denoted by $\rho_{q}^{m,N}$. There are no fields of the type $(C,A)$ in this subtheory.

Let us turn now to the heterotic string theory in four dimensions. We assume the diagonal modular
invariant for all the subtheories,
$$A_{\vec l,\vec l}=\prod_{i=1}^{n+r} \delta_{l_i,l_i}.\e$$
Again, we tensor $r$ minimal models, labeled by $k_i$, $i=1,2,\ldots, k_i$, along with $n$
nonstandard $N=2$ models, labeled by $(m_i,N_i)$, $i=1,2,\ldots,n$. We build the heterotic string theory
as described above, eq. (4.21), using the beta method and the heterotic map.
Since the $l$ indices are left right symmetric, the only $(C,C)$ fields in the internal theory are
those inherited from the chiral fields in each of the subtheories.
These fields are
$$D_{\vec p,\vec q}=\prod_{i=1}^r \Phi^{k_i}_{p_i} \prod_{i=1}^n \rho^{m_i,N_i}_{q_i},\e$$
where 
$$0\leq p_i\leq k_i,\  {\rm and\ } \qquad 0\leq q_i\leq N_i.\e$$ 
The generations in the theory come from $(C,C)$ fields
whose $U(1)$ charge is exactly $1$. These are the solutions of the equation
$$J=\sum_{i=1}^r J^{k_i}_{p_i}+\sum_{i=1}^n J^{m_i,N_i}_{q_i}=1.\e$$
Explicitly, this equation becomes,
$$\sum_{i=1}^{r} {p_i\over k_i+2}+\sum_{i=1}^n {N_i\over 2m_i-N} \left[ {m_i q_i\over N_i}\right]=1.\e$$
The generations in the string theory are in one to one correpondance with solutions of 
this equation for $p_i$ and $q_i$,  which obey eq. (5.7). 
We denote the solutions as $(\vec p,\vec q)$.
For the diagonal modular invariant, there are no additional 
generations. We emphasize that for non--diagonal modular invariants there may be additional generations.
The internal theory has the discrete symmetry group
$$G={\prod_{i=1}^r Z_{k_i+2} \times \prod_{i=1}^n Z_{2m_i-N_i}\over (g)},\e$$
where $(g)$ is the subgroup containing the $n+r$ long vectors $(\vec p,\vec q)$ such that
$\sum_i {p_i\over k_i+2}+\sum_i {m_i q_i\over 2m_i-N_i}\in Z$. 
The division by $(g)$ is due to the 
summation over $\beta$ in the partition function. When several of the subtheories are identical, the symmetry group, $G$, 
will
be enhanced by any permutation of the subtheories. 
According to eq. (3.31), the discrete symmetry charge of the field $D_{\vec p,\vec q}$ is given by
$$p_i\mod k_i+2,\qquad q_i+2 r_{q_i} \mod 2m_i-N_i,\e$$
where $r_q$, was defined as the integer,
$$r_q=\left[{mq\over N}\right]-{mq\over N}.\e$$

We turn now to the anti--generations. We define the lattice $\hat Q$ as the lattice spanned by the $n+r$ vector 
$(-1-,1,\ldots,-1,1,1,\ldots,1)$, where it is $-1$ for the first $r$ indices, 
which is the lattice inherited from $2Q$, where $Q$ is the lattice used in the beta method. 
We define
also the lattice $K$ spanned by the vectors $k_i+2$ on the $i$th model, and zero everywhere else, for $i=1,2,\ldots,r$, 
along with the vectors
$2m_j-N_j$ on the $j$th nonstandard model, $j=1,2,\ldots n$, and zero everywhere else.
The $(C,A)$ fields are obtained from left chiral field along with its right dagger, such that
$(\vec p,\vec q)$ obeys the total left $U(1)$ charge $1$ and the total right $U(1)$ charge $-1$. In addition, these 
fields must be obtainable by the beta method. This implies,
$$(\vec p,\vec q+2{\vec r}_q)\in \hat Q+K,\e$$
where ${\vec r}_q$ is the vector whose elements are $r_{q_i}$.
So this vector must lie in the combined lattice $\hat Q+K$. There are no additional antigenerations.
We conclude that the antigenerations are in one to one correspondence with solutions of the
two equations eqs. (5.9,5.13). The discrete symmetry charge of the anti--generations is $0$ for all the $Z_i$ groups.
For non--diagonal modular invariants there may be additional anti--generations.

Let us work out an example. Take $m=5$ and $N=7$. then the central charge of this theory is given by, eq. (3.25),
$$c={3 N\over 2 m-N}=7.\e$$
To get $c_{\rm int}=9$, needed to compactify to $4$ dimensions, we can tensor, along, two $k=1$ minimal
models, whose central charge is $1$. We denote this theory as $1^2(5,7)$.
The chiral fields of the $(5,7)$ subtheory are labeled by an integer $q$, $q=0,1,2,\ldots,7$.
Their $U(1)$ charges, according to eq. (2.31), are
$$J_q^{5,7}={N\over 2m-N} \left[{mq\over N}\right]=0,5/3,1,1/3,2,4/3,2/3,7/3,\e$$
for $q=0,1,2,3,4,5,6,7$, respectively. There are four solutions with total $U(1)$ charge $1$, solving eq. (5.9). 
These are given by
$$(p_1,p_2,q)=(0,0,2),\ (0,1,6),\ (1,0,6),\ (1,1,3).\e$$
We conclude that there are four generations in this model.

For the anti--generations, we need any of the above solutions, eq. (5.16), which lie in the $\hat Q+K$ lattice,
eq. (5.13). For this we calculate $q+2 r_q$. It is,
$q+2r_q=0,1,0,5,0,3,4,5\mod 6$, in the same order as above. With these values there is one solution to  
eq. (5.13), with the values of $(p_1,p_2,q)$ given in eq. (5.16). It is
$$(\vec p,\vec q)=(1,1,3).\e$$
We conclude that there is one
antigenerations in this model. The net number of generations, here, is thus three.
This means that the model $1^2(5,7)$ accurately predicts the number of generations found in nature.
\mysec{Discussion.}

We described here the construction of new parafermionic theories labeled by any two integers $m$ and $N$,
such that $m$ and $N$ are strange and $N/2< m< N$. The central charge of this rational theory is
$$c={4N-2 m\over 2m-N}.\e$$
We see that we get for the central charges any rational number above two, $c>2$.

We then tensor these new parafermions with a free boson to get new $N=2$ superconformal field theories.
This can be done for any $m$ and $N$. The central charge of the superconformal theory is
$\hat c=c+1$ or,
$$\hat c={3 N\over 2m-N}.\e$$
Again, we get any rational central charge above three for these theories, $\hat c>3$.

Next, we build string theories in even dimensions by tensoring several of the new superconformal models,
described here,
along with several minimal superconformal models. Following ref. \r\Gep, we get full fledged
superstring and heterotic strings in four dimensions. In four dimensions the internal theory has 
central charge $9$,
$$9=\sum_{i=1}^r {3k_i\over k_i+2}+\sum_{j=1}^n {3 N_j\over 2m_j-N_j}.\e$$
We see that we have an infinite number of solutions to this equation, and it follows that there is an 
infinite number of such string theories.

The heterotic theories in four dimensions have a gauge group which includes $E_8\times E_6$. The massless 
spectrum includes some chiral fermions in the representation $27$ of $E_6$ (generations) and some
chiral fermions in the $\bar{27}$ of $E_6$ (anti--generations). We gave a formula for the generations,
eq. (5.9), and antigenerations, eq. (5.13), along with their discrete symmetry charges, eq. (5.11).

It was conjectured in ref. \r\Gep\ that all the $N=2$ string theories correspond to compactification on some
Calabi--Yau manifold, originally discussed by Candelas et al. \r\Cand. 
We propose to find the manifolds corresponding to the new string theories discussed here.
The generations in the geometric 
picture are elements of the cohomology group $H^{2,1}$, whose number is $h^{2,1}$,
and the antigenerations are elements of the cohomology
$H^{1,1}$, whose number is $h^{1,1}$. Thus, we know the Euler number of the manifold $\chi=h^{2,1}-h^{1,1}$,
along with the discrete symmetry group and the representation of this discrete symmetry group on the 
elements of the cohomology. This provides ample information about the conjectured manifold.
At the present we do not know of examples of such manifolds, describing these new string theories, but we intend to 
further study this question in the future.

\ack
We thank S. Pal for his participation in the early stages of this work. We thank the Einstein center 
for support. Discussions with A. Belavin and A. Zamolodchikov are gratefully acknowledged.
D.G. would like to thank Ida Deichaite for constructive remarks on the manuscript. 

\refout

\bye